\documentclass[12pt]{iopart}
\usepackage{iopams}
\usepackage{graphicx}

\begin{document}

\newcommand{\ket}[1]{\left|#1\right>}
\newcommand{\bra}[1]{\left<#1\right|}
\newcommand{\sket}[1]{|#1\rangle}
\newcommand{\sbra}[1]{\langle#1|}
\newcommand{\sbraket}[2]{\langle#1|#2\rangle}
\newcommand{\sketbra}[2]{|#1\rangle\langle#2|}
\newcommand{\abs}[1]{\left|#1\right|}
\newcommand{\sabs}[1]{|#1|}
\newcommand{\norm}[1]{\left\|#1\right\|}
\newcommand{\snorm}[1]{\|#1\|}
\newcommand{\avr}[1]{\left<#1\right>}
\newcommand{\savr}[1]{\langle#1\rangle}
\newcommand{\s}[1]{^{(#1)}}

\title{Correlations in Quantum Spin Systems from the Boundary Effect}
\author{Jaeyoon Cho}
\address{School of Computational Sciences, Korea Institute for Advanced Study, Seoul 130-722, Korea}
\ead{choooir@gmail.com}
\vspace{10pt}
\begin{indented}\item[]\today
\end{indented}

\begin{abstract}
We introduce the boundary effect on the ground state as an attribute of general local spin systems that restricts the correlations in the ground state. To this end, we introduce what we call a boundary effect function, which characterises not only the boundary effect, but also the thermodynamic limit of the ground state. We prove various aspects of the boundary effect function to unfold its relationship to other attributes of the system such as a finite spectral gap above the ground state, two-point correlation functions, and entanglement entropies. In particular, it is proven that an exponentially decaying boundary effect function implies the exponential clustering of two-point correlation functions in arbitrary spatial dimension, the entanglement area law in one dimension, and the logarithmically corrected area law in higher dimension. It is also proven that gapped local spin systems with nondegenerate ground states ordinarily fall into that class. In one dimension, the area law can also result from a moderately decaying boundary effect function, in which case the system is thermodynamically gapless.
\end{abstract}
\maketitle


\section{Introduction}

One of the prominent approaches to quantum many-body theory is to explore universal features of typical classes of many-body systems, thereby guiding our intuition into more specific problems. Commonly studied in that context are spin systems with finite-range interaction. The {\em localness} of the interaction is then manifested as various dynamical and static features. A quintessential example is the Lieb-Robinson bound, which demonstrates that local interaction manifests itself as a locality in the emergent dynamics~\cite{lieb72,nach06}.

Regarding the static features, various aspects of correlation that ground states exhibit are of primary concern. In particular, along with the condition of local interaction, the existence of a finite spectral gap above the ground state highly restricts the correlation that the ground state can accommodate. In such systems, arbitrary two-point correlation functions in the ground state decay exponentially with distance, called the exponential clustering theorem~\cite{nach06,hastings06}. Moreover, such ground states can accommodate only a restricted amount of entanglement. When the entanglement between a subregion and the rest, called the entanglement entropy, scales at most as the boundary size of the region, such a state is said to obey the entanglement area law~\cite{eisert10}. It turns out that in one-dimensional gapped spin systems with local interaction, nondegenerate ground states obey the area law~\cite{hastings07}. This contrasts with the case of random states, which exhibit an entanglement entropy proportional to the volume of the region~\cite{page93}. Conceptually, this implies that the ground states obeying the area law occupy only an extremely small portion of the Hilbert space, which again suggests that we do not actually need too many parameters to describe them. The area law is thus of crucial importance both conceptually and practically, e.g., in the area of the simulation of quantum many-body systems~\cite{scho11}, topological quantum phases~\cite{kitaev06,levin06},  and the Hamiltonian complexity theory~\cite{gha14}. For this reason, a general proof of the area law in more than one dimension has been awaited for quite a while~\cite{eisert10,hastings07,aud02,plenio05,wolf06,wolf08,masanes09,beau10,gott10,aharonov11,arad12,mic12,mic13,aco13,bra13}. Very recently, the author has given such a proof under reasonable sufficient conditions~\cite{cho14}.

Apparently, all the attributes mentioned above---local interaction, spectral gap, spatial dimension, Lieb-Robinson bound, exponential clustering, area law, etc.---are deeply related to each other and inseparable. However, our understanding of them and their relationship is far from being satisfactory. Owing to the inherent complexity, the ordinary task is rather to figure out various aspects of their relationship. For example, although one can infer various features assuming the existence of a finite spectral gap, it is extremely hard to know whether a given Hamiltonian is gapped or not. 

In this paper, we bring in the boundary effect on the ground state, in a specific sense we clarify later, as yet another attribute of many-body systems that strongly dictates the ground state correlations. The boundary effect is certainly a widespread concept throughout the whole area of physics, although its actual manifestation may vary \textit{ad hoc} from problem to problem. To name just a few examples in the context of many-body physics, integrable systems with open boundary conditions, amenable to analytical treatments, have provided a rich playground to quantitatively study boundary-originated changes in one dimensional systems~\cite{skl88,gho94} and the boundary (edge) mode in topologically ordered systems has been one of the prime issues in modern condensed matter physics~\cite{wen91,has10}. In most cases, it is presumably safe to say that the boundary effect is, as such, any change to the physical properties of the system owing to the presence of a boundary. This concept is meaningful only when there are corresponding physical properties in the absence of the boundary, which will be those of the infinite (unbounded) system, i.e., those in the thermodynamic limit. Empirically, we expect, albeit not necessarily true in general, that the boundary effect is somehow localised near the boundary in ordinary many-body systems without altering the bulk properties, as in, e.g., topologically ordered systems wherein the edge mode appears to be spatially localised at the boundary while the bulk retains the properties in the thermodynamic limit~\cite{wen91,has10}. However, this intuition seems rather unrefined. How can we single out the boundary effect for a given general many-body system? How can we quantify the degree to which the boundary effect permeates into the bulk? One of our aims in this paper is to suggest one possible way of answering those questions in terms of the ground state wavefunction, although it can be generalised, e.g., to the cases of thermal states. It will turn out that when the boundary effect on the ground state is localised at the boundary, the ground state correlations (in the bulk) are  restricted thereby, and that a finite spectral gap above the ground state plays a role of localising the boundary effect. This observation allows us to offer a sensible and intuitive picture on how local nature can emerge in the ground states.

As was discussed above, in order to address the boundary effect, we consider a thermodynamic limit of the ground state as a reference state in the absence of the boundary, which is then compared to that in the presence of the boundary so that their difference can be identified with the boundary effect. This is done by defining what we call a boundary effect function, which essentially characterises the change made to the ground state when the system is enlarged. It turns out that the boundary effect function characterises the convergence of the ground state towards its thermodynamic limit and also how far the boundary effect permeates into the bulk. We prove various aspects of the boundary effect function. First, the boundary effect function decays exponentially in gapped spin systems except for an unusual exception. Second, the boundary effect function gives a bound to arbitrary two-point correlation functions in the ground state. In particular, if the boundary effect function decays exponentially, all two-point correlation functions decay exponentially with the distance in any spatial dimension, i.e., the ground state obeys the exponential clustering. Third, the boundary effect function gives a bound to entanglement entropies of the ground state. In particular, if the boundary effect function decays exponentially, the entanglement entropy obeys the area law in one dimension and the logarithmically-corrected area law in higher dimension. Remarkably, in one dimension, the area law can also result from a moderately decaying boundary effect function, in which case the system should be thermodynamically gapless. This may be related to the remarkable performance of the density matrix renormalisation group method in one dimension regardless of the gapness of the system~\cite{scho11}. 

\section{Notations and Assumptions}

We will follow the notations used in \cite{cho14}. We consider systems of finite-dimensional spins residing on a $D$ dimensional lattice with one spin per site. As will be discussed later, it is essential in our discussion to vary the number of spins in the system and investigate how it alters the system. As such,  a spin system in our discussion formally means a set of different-sized systems having the same microscopic natures (for example, imagine a case where we consider a certain type of Heisenberg chain and take its thermodynamic limit). This scenario covers almost all physical problems we ordinarily face. We are interested in the cases where the interactions between spins are local. For the notion of localness to make sense, we make two assumptions on the lattice. First, there is a constant $a_0$ such that $\ell_E(s,s')\le a_0 \ell_G(s,s')$ for any sites $s$ and $s'$, where $\ell_E(s,s')$ and $\ell_G(s,s')$ are the Euclidean distance and the graph distance, respectively. Second, one can take a unit volume $(\delta l)^D$ such that the number of sites in a unit volume is bounded by $n_0(\delta l)^D$ for some constant $n_0$. Note that these two properties are very general. For convenience, let us define for given site $s$, the set of neighbouring sites
\begin{equation}
\mathcal{B}_s^k=\{\mathrm{site~}s':\ell_G(s,s')<k\}
\end{equation}
and denote by $\mathcal{U}(\mathcal{B}_s^k)$ the set of unitary operators supported on $\mathcal{B}_s^k$. As we consider local interactions, the $n$-spin Hamiltonian of the system can be written as
\begin{equation}
H\s{n}=\sum_{s=1}^n h_s\s{n}=H\s{n-1}+K_n
\end{equation}
without loss of generality, where $h_s\s{n}$ is supported on $\mathcal{B}_s^{k_0}$ with $k_0$ being a constant bounding the range of the interaction. Following common practice, we index the spins in such a way that the $n$-spin system is constructed by adding the $n$th spin on the boundary of the $(n-1)$-spin system, though this rule can be tailored for specific problems at hand. Note that due to the boundary terms, $K_n$ is in general supported on $\mathcal{B}_n^{2k_0}$, not $\mathcal{B}_n^{k_0}$, and $h_s\s{n}=h_s\s{n-1}$ for $s\not\in\mathcal{B}_n^{k_0}$\footnote{For example, imagine a one-dimensional chain with three-body interaction terms in the bulk and two-body terms at the boundary, for which $k_0=2$. In this case, $K_n=h_{n-1}\s{n}+h_n\s{n}-h_{n-1}\s{n-1}$ as $h_{n-1}\s{n}\not=h_{n-1}\s{n-1}$.}. We assume that $H\s{n}$ has a {\em nondegenerate} ground state, which we denote by $\sket{\Psi_0\s{n}}$. For an operator $A$, we use both the trace norm $\norm{A}_1=\Tr \sabs{A}$ and the operator norm $\norm{A}_\infty$ that is the largest eigenvalue of $\sabs{A}$. For a vector $\sket\psi$, $\snorm{\sket\psi}$ denotes the Euclidean norm $\sqrt{\langle\psi|\psi\rangle}$. We assume that $J=\sup_{s,n}\snorm{h_s\s{n}}_\infty$ is finite.

\section{Boundary Effect Function}

A boundary effect can be defined as any difference between the properties of the system in the presence and the absence of the boundary. In the absence of the boundary, the material properties are those in the thermodynamic limit. As such, the boundary effect can be meaningfully discriminated from other effects only when the system has a thermodynamic limit. We define the existence of a thermodynamic limit for the ground state as follows. Suppose we take a subregion $A$ of the system and obtain the reduced density matrix $\rho_A\s{n}=\Tr_{\setminus A}\sket{\Psi_0\s{n}}\sbra{\Psi_0\s{n}}$ of the ground state $\sket{\Psi_0\s{n}}$, where $\setminus A$ denotes the complement of region $A$. For a thermodynamic limit to exist, all the local observables on region $A$ should become intensive quantities by losing their dependency on the system size in the large $n$ limit. This is possible only when the reduced density matrix converges as 
\begin{equation}
\rho_A=\lim_{n\rightarrow\infty}\rho_A\s{n}. 
\end{equation}
When this is the case, we will say that the ground state has a thermodynamic limit. Note that we are considering open boundary conditions, by construction.

\begin{figure}
\begin{tabular}{cc}
\includegraphics[width=0.46\textwidth]{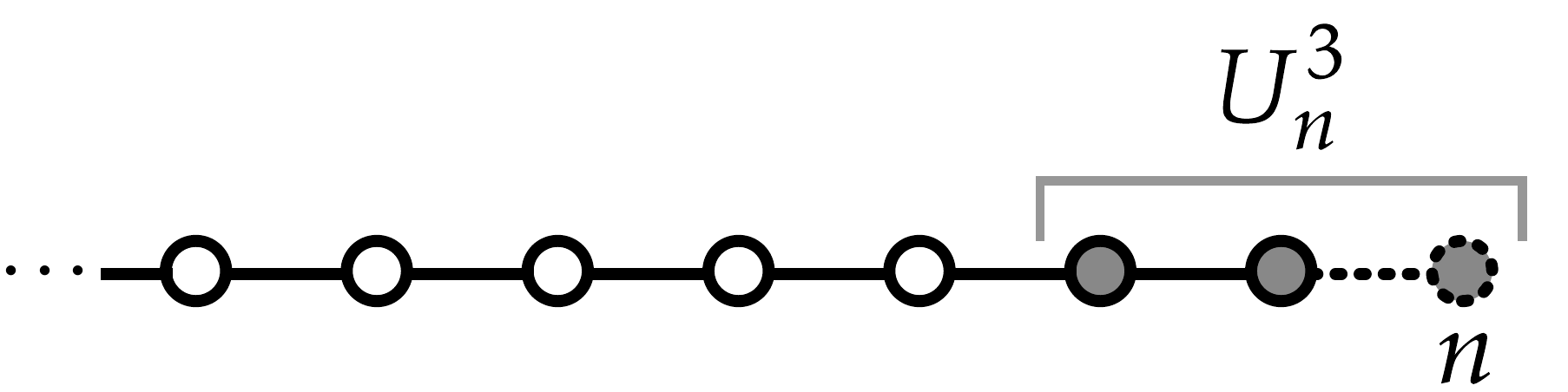} & 
\includegraphics[width=0.46\textwidth]{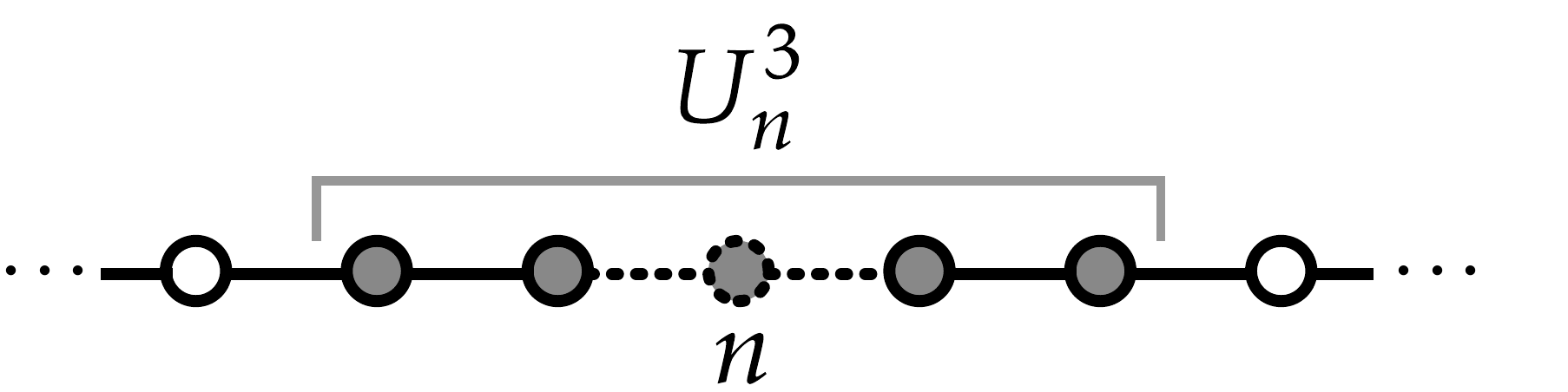} \\
(a) & (b)
\end{tabular}
\caption{(a) The spin chain of length $(n-1)$ is enlarged by adding the $n$th spin. (b) Two independent spin chains are combined by the $n$th spin into one larger spin chain. For illustrative purpose, the spins on which $U_n^3$ acts are depicted by shaded circles.}
\label{fig:befunction}
\end{figure}

In order to quantify the boundary effect, consider a mapping from $\sket{\Psi_0\s{n-1}}$ to $\sket{\Psi_0\s{n}}$. We are willing to find an optimal unitary transformation $U$ such that $U\sket{\Psi_0\s{n-1}}\ket{0}_n$ best approximates $\sket{\Psi_0\s{n}}$, where $\ket{0}_n$ is an arbitrary state of spin $n$. As $H\s{n-1}$ and $H\s{n}$ differ only by a local term $K_n$ supported on $\mathcal{B}_n^{2k_0}$, it is reasonable to define for $r>0$ an optimal unitary transformation $U_n^r\in\mathcal{U}(\mathcal{B}_n^r)$ such that
\begin{eqnarray}
\mu_n(r)&=\frac{1}{\sqrt{2}}\norm{\sket{\Psi_0\s{n}}-U_n^r\sket{\Psi_0\s{n-1}}\sket{0}_n}\nonumber\\
&=\frac{1}{\sqrt{2}}\inf_{U\in\mathcal{U}(\mathcal{B}_n^r)}\norm{\sket{\Psi_0\s{n}}-U\sket{\Psi_0\s{n-1}}\sket{0}_n}.
\end{eqnarray}
Note that
\begin{equation}
\mu_n(r)=\sqrt{1-\abs{\sbra{\Psi_0\s{n}}U_n^r\sket{\Psi_0\s{n-1}}\sket{0}_n}}\le 1.
\end{equation}
Apparently, $\mu_n(r)$ is a non-increasing function of $r$ and
\begin{equation}
\lim_{r\rightarrow\infty}\mu_n(r)=0.
\end{equation}
Let us further define a non-increasing function
\begin{equation}
\mu(r)=\sup_n\mu_n(r),
\end{equation}
which we will call a boundary effect function. Here, $\mu_n(r)$, hence $\mu(r)$, may depend on how the spins are indexed. For example, in one dimension, one can consider two different cases as shown in Fig.~\ref{fig:befunction}.

It is remarkable that the boundary effect function characterises the convergence of the ground state towards the thermodynamic limit. For example, consider a one-dimensional chain of $n$ spins and let region $A$ be the sites from 1 to $m$. If a thermodynamic limit exists, one can take a Cauchy sequence such that $\lim_{n\rightarrow\infty}\eta_{A}\s{n}=0$ with 
\begin{equation}
\eta_{A}\s{n}=\frac{1}{2}\norm{\rho_A\s{n}-\rho_{A}\s{n-1}}_1.
\end{equation}
Note that 
\begin{equation}
\mu(n-m)^{2}\le\eta_{A}\s{n}\le\sqrt{2\mu(n-m)^2-\mu(n-m)^4},
\end{equation} 
which follows from the Uhlmann's theorem and the inequalities between the trace distance and the fidelity~\cite{nielsen}. The boundary effect function thus characterises how fast the system approaches a thermodynamic limit as $n$ is increased. From a different perspective, it also characterises how far the boundary effect permeates into the bulk. For example, if $\mu(r)=0$ for $r>r_0$ and $n>m+r_0$, the region $A$ cannot recognise the existence of a boundary and $\rho_A\s{n}$ becomes independent of $n$. 

In general, the boundary effect function would be a certain decreasing function, depending on the  system under consideration. For one dimensional spin chains, its meaning is clearer as the ground state has a thermodynamic limit if and only if the boundary effect function asymptotically vanishes. In higher dimension, however, the boundary effect function as defined above does not fully characterise the boundary effect since $\mu(r)$ encapsulates the influence of only a tiny change in the entire boundary region. In higher dimension, while the ground state has a thermodynamic limit if the boundary effect function asymptotically vanishes, the inverse is not necessarily true as one can imagine an exceptional situation where the boundary effect function does not vanish even though the reduced density matrix for the bulk region converges (e.g., if an addition of a spin exclusively alters only the entire boundary region). It is unclear, however, if such an exceptional case is possible. 

The functional form of the boundary effect function, which characterises the convergence rate toward the thermodynamic limit as explained above, is also important. If somehow $\mu(r)$ is nonzero for $r\le r_0$ for some constant $r_0$ and has an exponentially decreasing tail outside, we may argue that the boundary effect is quasi-local. In this case, we can call $U_n^\infty$ a quasi-local unitary transformation as it can be approximated well by a local unitary transformation. Moreover, we can call the mapping $\sket{\Psi_0\s{n}}\rightarrow\sket{\Psi_0\s{n+1}}\rightarrow\sket{\Psi_0\s{n+2}}\rightarrow\cdots$ a quasi-local extension of the ground state in the sense that the size is enlarged by a series of quasi-local unitary transformations.

\section{Boundary Effect Function in Gapped Systems}

In the previous section, the boundary effect function was defined as a characteristic function of a system, independent of other attributes of the system. An immediately following question is then how the boundary effect function is determined. In this section, we prove that a finite spectral gap above the ground state restricts the boundary effect function. To be specific, we prove that gapped systems have an exponentially decaying boundary effect function unless $\mu(r)=1$ for all $r$ (in this exceptional situation, an addition of a single spin can drive a quantum phase transition of the entire system). As discussed in the previous section, systems with exponentially decaying boundary effect functions have a quasi-locally extended ground state. In the ensuing sections, we will see that such a local nature of the ground state is indeed manifested as restricted correlations in the ground state.

Suppose the system has a finite spectral gap larger than $\Delta$ between the ground state and the first excited state. In order to consider a mapping from $\sket{\Psi_0\s{n-1}}\ket{0}_n$ to $\sket{\Psi_0\s{n}}$, let us consider a Hamiltonian
\begin{equation}
\tilde{H}\s{n-1}=H\s{n-1}+\Delta(I-\ket{0}_n\!\bra{0}),
\end{equation}
where $I$ is the identity operator. It is clear that this Hamiltonian preserves the gap condition and its ground state is $\sket{\tilde\Psi_0\s{n-1}}=\sket{\Psi_0\s{n-1}}\ket{0}_n$. Let us define the reduced density matrix
\begin{equation}
\rho_{\setminus s,k}\s{n}=\Tr_{\mathcal{B}_s^k}\sket{\Psi_0\s{n}}\sbra{\Psi_0\s{n}}.
\end{equation}
If $\mu(r)$ is not constantly one, there is a constant $l_0$ such that $\mu(l_0)<1$ and hence
\begin{equation}
\frac{1}{2}\norm{\rho_{\setminus n,l_0}\s{n}-\rho_{\setminus n,l_0}\s{n-1}}\le\sqrt{2\mu(l_0)^2-\mu(l_0)^4}<1.
\end{equation}
 For later convenience, let $K_0$ be the sum of the common terms appearing both in $\tilde{H}\s{n-1}$ and $H\s{n}$ supported on the complement of $\mathcal{B}_n^{l_0}$ and let
 \begin{eqnarray}
U_n^{l_0}\tilde{H}\s{n-1}U_n^{l_0\dagger}=K_0+K_1,\\
H\s{n}=K_0+K_2.
 \end{eqnarray}
 It then follows from the lemma in \cite{cho14} that one can introduce an ancillary two-level system $a$  to construct a local Hamiltonian for an adiabatic passage with a single parameter $\lambda(t)\in[0,1]$:
\begin{eqnarray}
H(\lambda(t))&=H_0+V(\lambda(t))\nonumber\\
&=K_0+(K_1+\Delta)\otimes\ket{1}_a\!\bra{1}+K_2\otimes\ket{2}_a\!\bra{2}+V(\lambda(t)),
\end{eqnarray}
where $V(\lambda)$ is a local term supported on $\mathcal{B}_n^{l_0+k_0}$ plus the ancillary system $a$ and
\begin{eqnarray}
V(0)=-\Delta\ket{1}_a\!\bra{1}+\Delta\ket{2}_a\!\bra{2},\\
V(1)=0
\end{eqnarray}
so that the ground state of $H(0)$ is $\ket{\psi_i}=U_n^{l_0}\sket{\tilde{\Psi}_0\s{n-1}}\ket{1}_a$ and that of $H(1)$ is $\ket{\psi_f}=\sket{\Psi_0\s{n}}\ket{2}_a$. Here, the spectral gap of $H(\lambda)$ is at least larger than $\Delta_a=\mu(l_0)^2\Delta/10$ for all $\lambda$~\cite{cho14}. The intuition behind the proof is that the adiabatic passage can be performed in a finite time scale of $\mathcal{O}(1/\Delta_a)$ by changing the Hamiltonian only locally and hence, owing to the Lieb-Robinson bound, this local change can affect the system only quasi-locally, causing the boundary effect function to decay exponentially. This idea can be formulated as follows.

Let $T$ be the time for the adiabatic passage, i.e., $\lambda(0)=0$ and $\lambda(T)=1$. Let
\begin{equation}
\sket{\psi(t)}=\mathcal{T}\left[\exp\left\{-i\int_{0}^{t}H(\lambda(t'))dt'\right\}\right]\ket{\psi_i},
\end{equation}
where $\mathcal{T}$ is the time-ordering operator. In the interaction picture taking $\ket{\psi(t)}_I=e^{iH_0t}\ket{\psi(t)}$,
\begin{equation}
\sket{\psi(T)}_{I}= e^{iH_0T}\sket{\psi(T)}=U_{I}(T)\sket{\psi_i},
\end{equation}
where
\begin{equation}
U_{I}(T)=\mathcal{T}\left[\exp\left\{-i\int_{0}^{T}V_I(t)dt\right\}\right]
\end{equation}
with $V_I(t)=e^{iH_0t}V(\lambda(t))e^{-iH_0t}$. Exploiting the adiabatic theorem in Ref.~\cite{lidar09},
\begin{eqnarray}
\epsilon_{a}(T)&=\inf_{\theta}\norm{\ket{\psi_f}-e^{i\theta}\sket{\psi(T)}_{I}}\nonumber\\
&=\inf_{\theta}\norm{e^{iH_0T}\left(\sket{\psi_f}-e^{i\theta}\sket{\psi(T)}\right)}\nonumber\\
&\le\mathcal{O}\left[T^{1+\gamma}\exp\left(-\frac{c\gamma\Delta_{a}^{3}T}{J^{2}}\right)\right],
\end{eqnarray}
 where $c$ and $\gamma$ are positive constants of $\mathcal{O}(1)$. Note that there always exist positive constants $T_{0}$ and $\tilde{c}<c\gamma$ such that
\begin{equation}
\epsilon_{a}(T)\le\mathcal{O}\left[\exp\left(-\frac{\tilde{c}\Delta_{a}^{3}T}{J^{2}}\right)\right]\mathrm{~for~}T\ge T_{0}. 
\end{equation}
As we are interested in the limit of large $T$, we take this regime to simplify the formula. We also need to bound 
\begin{equation}
\epsilon_{b}(T)=\inf_{U\in\mathcal{U}(\mathcal{B}_{n}^{r})}\norm{\sket{\psi(T)}_{I}-U\sket{\psi_i}} 
\end{equation}
so that 
\begin{equation}
\mu(r)\le\frac{1}{\sqrt{2}}\left[\epsilon_{a}(T)+\epsilon_{b}(T)\right].
\end{equation}
Let $\sigma=\sket{\psi_i}\sbra{\psi_i}$. It follows that
\begin{eqnarray}
\epsilon_{b}(T)&\le2\norm{\Tr _{\mathcal{B}_n^r}\{U_{I}(T)\sigma U_{I}^{\dagger}(T)\}-\Tr _{\mathcal{B}_n^r}\sigma}_{1}\nonumber\\
&=4\Tr _{\setminus\mathcal{B}_n^r}\left[\Lambda_{\setminus n,r}\Tr _{\mathcal{B}_n^r}\{U_{I}(T)\sigma U_{I}^{\dagger}(T)-\sigma\}\right],
\end{eqnarray}
where $0\le\Lambda_{\setminus n,r}\le I$ supported on the complement of $\mathcal{B}_n^r$ is an arbitrary operator maximizing the bound. We thus have
\begin{eqnarray}
\epsilon_{b}(T)&\le4\Tr \left[\Lambda_{\setminus n,r} U_{I}(T)\sigma U_{I}^{\dagger}(T)-\Lambda_{\setminus n,r}\sigma\right]\nonumber\\
&=4\Tr \left[\Lambda_{\setminus n,r} U_{I}(T)\sigma U_{I}^{\dagger}(T)-U_{I}(T)\Lambda_{\setminus n,r}\sigma U_{I}^{\dagger}(T)\right]\nonumber\\
&\le4\norm{[\Lambda_{\setminus n,r},U_{I}(T)]\sigma}_{1}\le4\norm{[\Lambda_{\setminus n,r},U_{I}(T)]}_{\infty}.
\end{eqnarray}
Exploiting the Lieb-Robinson bound
\begin{equation}
\norm{[\Lambda_{\setminus n,r},V_I(t)]}_{\infty}\le\mathcal{O}\left(\norm{\Lambda_{\setminus n,r}}_{\infty}\norm{V(\lambda(t))}_{\infty}e^{\upsilon t-\xi r}\right),
\end{equation}
 where $\upsilon$ and $\xi$ are positive constants, letting $V_\infty=\snorm{V(\lambda(t))}_{\infty}=\mathcal{O}[J(l_0+k_0)^D]$, and noting that
\begin{equation}
U_{I}(T)=1+\mathcal{T}\left\{\sum_{m=1}^{\infty}\frac{(-i)^{m}}{m!}\int_{0}^{T}dt_{1}\cdots dt_{m}V_I(t_{1})\cdots V_I(t_{m})\right\},
\end{equation}
we finally have
\begin{eqnarray}
\epsilon_{b}(T)&\le\mathcal{O}\left(\sum_{m=1}^{\infty}\frac{m}{m!}\int_{0}^{T}dt\,e^{\upsilon t-\xi r}V_{\infty}^{m-1}T^{m-1}\right)\nonumber\\
&\le\mathcal{O}\left[e^{(\upsilon+V_{\infty})T}e^{-\xi r}\right],
\end{eqnarray}
where we have used the commutator identity
\begin{equation}
\fl [A,B_1B_2\cdots B_N]=[A,B_1]B_2\cdots B_N+B_1[A,B_2]\cdots B_N+\cdots+B_1B_2\cdots[A,B_N].
\end{equation}
By taking
\begin{equation}
T=\frac{\xi}{\upsilon+V_{\infty}+\tilde{c}\Delta_{a}^{3}/J^{2}}r,
\end{equation}
we end up with an exponentially decaying boundary effect function
\begin{equation}
\mu(r)\le\mathcal{O}\left[\exp\left(-\frac{\xi\tilde{c}\Delta_{a}^{3}}{J^{2}(\upsilon+V_{\infty})+\tilde{c}\Delta_{a}^{3}}r\right)\right].
\end{equation}

\section{Correlation Functions in the Ground State}

\begin{figure}
\begin{tabular}{cc}
\includegraphics[width=0.46\textwidth]{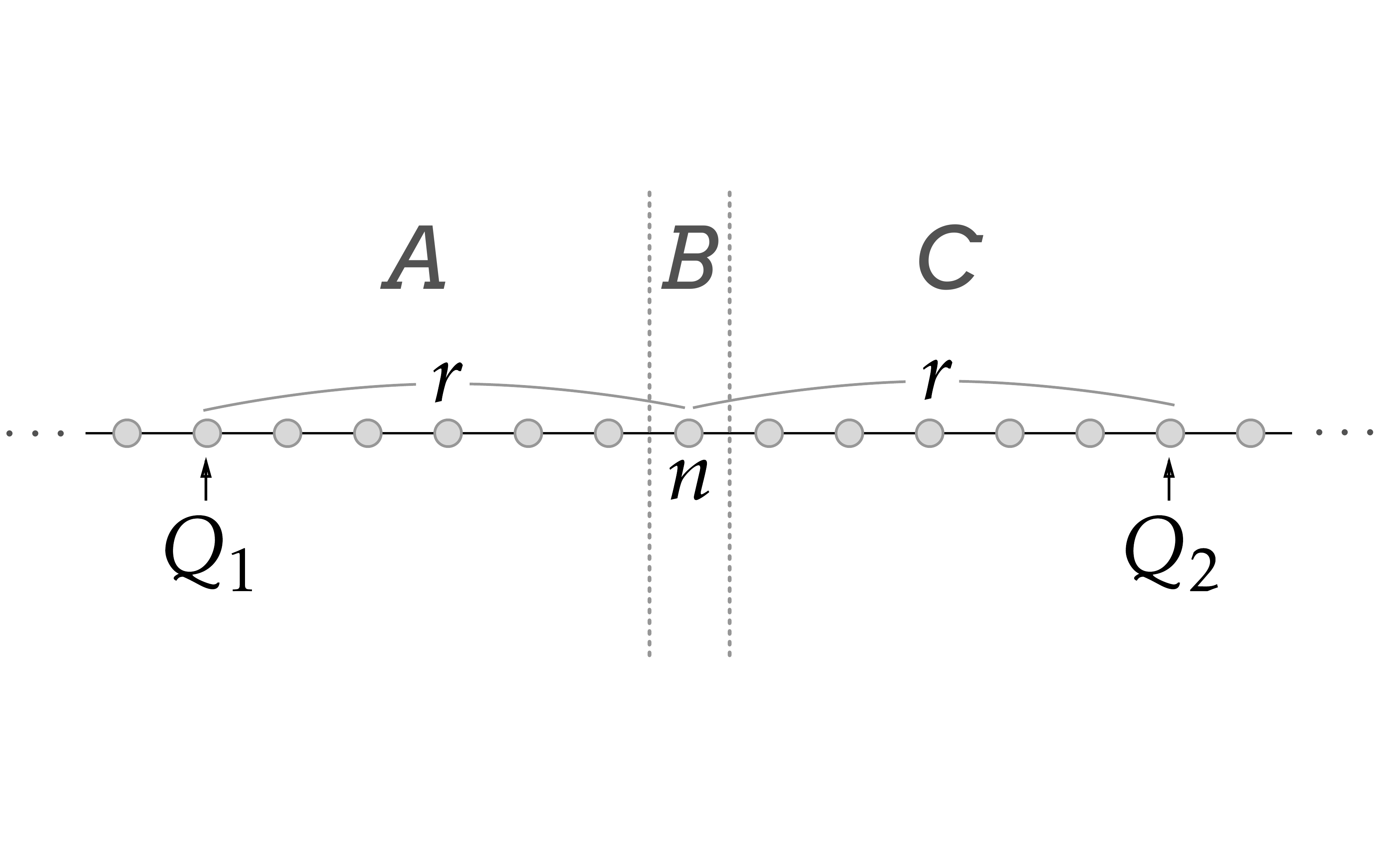} & 
\includegraphics[width=0.46\textwidth]{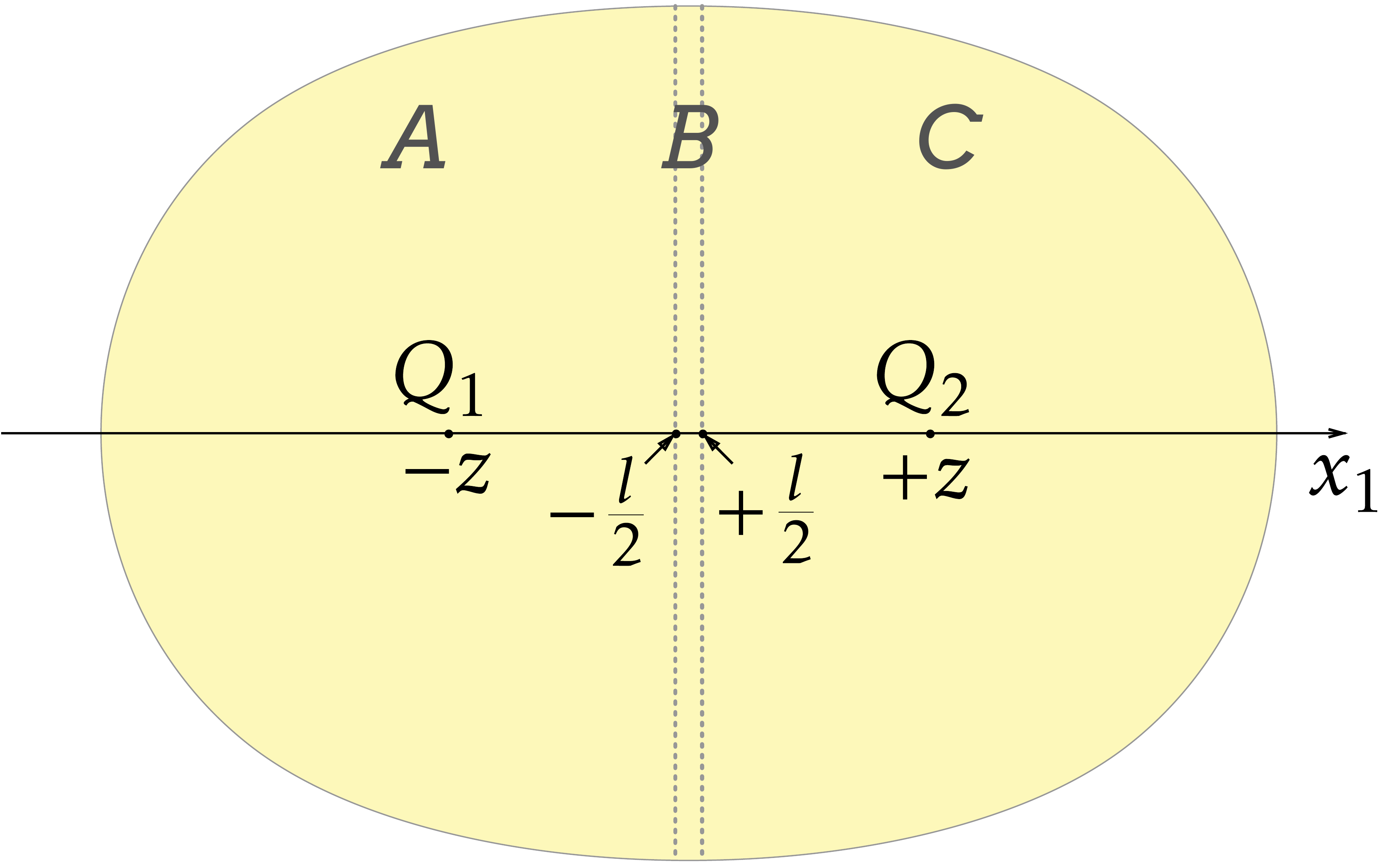} \\
(a) & (b)
\end{tabular}
\caption{Division of the system to bound two-point correlation functions (a) in one dimension and (b) in higher dimension.}
\label{fig:cfunction}
\end{figure}

In this section, we show that the boundary effect function gives a bound to arbitrary two-point correlations functions in the ground state. Let us first consider a one-dimensional chain of $n$ spins. We divide the chain into three regions as shown in Fig.~\ref{fig:cfunction}(a). We start from a system of $n-1$ spins composed of two independent spin chains in region $A$ and region $C$. The ground state of this system can be written as $\sket{\Psi_0\s{n-1}}=\sket{\Psi_0}_A\sket{\Psi_0}_C$. We then add a spin in region $B$ to get the desired $n$-spin system. The $n$-spin ground state can be written as
\begin{equation}
\sket{\Psi_0\s{n}}=U_n^\infty\sket{\Psi_0}_A\sket{0}_n\sket{\Psi_0}_C.
\end{equation}
Our aim is to bound the two-point correlation function
\begin{equation}
f(Q_{1},Q_{2})=\avr{Q_{1}Q_{2}}-\avr{Q_{1}}\avr{Q_{2}},
\end{equation}
where $Q_{1}$ and $Q_{2}$ are arbitrary single-spin operators supported on the $r$th neighbouring sites from site $n$, respectively, as shown in Fig.~\ref{fig:cfunction}(a). Our strategy is to approximate $\sket{\Psi_{0}\s{n}}$ with
\begin{equation}
\sket{{\Psi_{0}'}\s{n}}= U_{n}^{r}\ket{\Psi_{0}}_{A}\ket{0}_{n}\ket{\Psi_{0}}_{C}.
\end{equation}
Let us denote by $\avr{\cdot}'$ the average over $\sket{{\Psi_{0}'}\s{n}}$. Note that
\begin{equation}
\avr{Q_{1}Q_{2}}'-\avr{Q_{1}}'\avr{Q_{2}}'=0.
\end{equation} 
We use two inequalities. First, for any states $\sket{\psi}$ and $\sket{\phi}$~\cite{lidar08}, 
\begin{eqnarray}
\abs{\sbra{\psi}X\sket{\psi}-\sbra{\phi}X\sket{\phi}}&=\abs{\Tr [(\sketbra{\psi}{\psi}-\sketbra{\phi}{\phi})X]}\nonumber \\
&\le\norm{\sketbra{\psi}{\psi}-\sketbra{\phi}{\phi}}_1\norm{X}_\infty.
\end{eqnarray}
Second, 
\begin{equation}
\norm{\sketbra{\psi}{\psi}-\sketbra{\phi}{\phi}}_1\le2\inf_{\theta}\norm{\sket\psi-e^{i\theta}\sket\phi},
\end{equation} 
which follows from the inequality between the trace distance and the fidelity. From these, we have 
\begin{eqnarray}
\abs{f(Q_1,Q_2)}&=\left|\avr{Q_1Q_2}-\avr{Q_1Q_2}'+\avr{Q_1}'\left(\avr{Q_2}'-\avr{Q_2}\right)\right.\nonumber \\
&\quad\left.+\left(\avr{Q_1}'-\avr{Q_1}\right)\avr{Q_2}\right|\nonumber \\
&\le 6\sqrt{2}\norm{Q_1}_\infty\norm{Q_2}_\infty\mu(r).
\end{eqnarray}
The boundary effect function thus directly bounds the two-point correlation function. Note that quasi-locally extended states obey the exponential clustering. 

The same logic applies in higher dimension as well. For $D\ge2$, we take a near-continuum limit for analytical simplicity. As shown in Fig.~\ref{fig:cfunction}(b), we take the Cartesian coordinates $(x_{1},\cdots,x_{D})$ and divide the regions so that region $B$ is a $(D-1)$-dimensional plane with thickness $l$ defined by $l/2\le x_{1}\le l/2$ and region $A$ ($C$) is the region $x_{1}<l/2$ ($x_{1}>l/2$). We again bound the two-point correlation function $f(Q_{1},Q_{2})$, where $Q_{1}$ and $Q_{2}$ are, respectively, supported on the lattice sites at $\vec{x}_{\pm}=(\pm z,0,\cdots,0)$. Suppose there are $m$ spins in region $A+C$. We start from the state $\sket{\Psi_{0}\s{m}}=\ket{\Psi_{0}}_A\ket{\Psi_{0}}_C$ and extend it to $\sket{\Psi_{0}\s{n}}=\sket{\Psi_{0}}_{{ABC}}$ by filling up the region $B$. 
For brevity, let
\begin{equation}
\sket{\tilde{\Psi}_0\s{s}}=\sket{\Psi_0\s{s}}\ket{0}_{s+1}\cdots\ket{0}_n.
\end{equation} 
Note that
\begin{equation}
\sket{\Psi_0\s{n}}=U_n^\infty\sket{\tilde\Psi_0\s{n-1}}=U_n^\infty U_{n-1}^\infty\sket{\tilde\Psi_0\s{n-2}}=U_{n}^\infty U_{n-1}^\infty\cdots U_{m+1}^\infty\sket{\tilde\Psi_0\s{m}}. 
\end{equation}
We approximate $\sket{\Psi_0\s{n}}$ with
\begin{equation}
\sket{{\Psi_0'}\s{n}}= U_{n}^{r_n} U_{n-1}^{r_{n-1}}\cdots U_{m+1}^{r_{m+1}}\sket{\tilde\Psi_0\s{m}},
\end{equation} 
where $r_s=\min\{\ell_{G}(s,\vec{x}_+),\ell_{G}(s,\vec{x}_-)\}$. Note that $U_s^{r_s}$ commutes with $Q_1$ and $Q_2$ for all $s$ and thus we can use the same technique as in the one-dimensional case. From the inequality
\begin{eqnarray}
&\norm{U_n^\infty\cdots U_{m+1}^\infty\sket{\tilde\Psi_0\s{m}}-U_n^{r_n}\cdots U_{m+1}^{r_{m+1}}\sket{\tilde\Psi_0\s{m}}}\nonumber\\
&\le\norm{(U_n^\infty-U_n^{r_n})\sket{\tilde\Psi_0\s{n-1}}}+\norm{U_{n}^{r_{n}}(U_{n-1}^\infty-U_{n-1}^{r_{n-1}})\sket{\tilde\Psi_0\s{n-2}}}+\nonumber\\
&\quad\cdots+\norm{U_{n}^{r_{n}}\cdots U_{m+2}^{r_{m+2}}(U_{m+1}^\infty-U_{m+1}^{r_{m+1}})\sket{\tilde\Psi_0\s{m}}}\\
&\le\sqrt{2}\sum_{s=m+1}^n\mu(r_s),
\end{eqnarray}
it follows that
\begin{equation}
\sabs{f(Q_1,Q_2)}\le6\sqrt{2}\norm{Q_1}_\infty\norm{Q_2}_\infty\sum_{s=m+1}^n\mu(r_s).
\end{equation}

Now suppose the boundary effect function decreases exponentially as $\mu(r)\le\mu_0e^{-\kappa r}$.
For $r\gg l$,
\begin{eqnarray}
\sum_{s=m+1}^{n}\mu(r_s)&\le\mu_{0}n_{0}l\int_{-\infty}^\infty dx_{2}\cdots dx_{D}e^{-\frac{\kappa}{a_{0}}\sqrt{z^2+x_{2}^2+\cdots+x_{D}^{2}}}\nonumber\\
&=\mu_{0}n_{0}lz^{D-1}\int_{-\infty}^{\infty} dy_{2}\cdots dy_{D}e^{-\frac{\kappa z}{a_{0}}\sqrt{1+y_{2}^2+\cdots+y_{D}^{2}}}\nonumber\\
&=\mathcal{O}\left[z^{D-1}\int_{0}^{\infty} dy\,y^{D-2}e^{-\frac{\kappa z}{a_{0}}\sqrt{1+y^{2}}}\right]\nonumber\\
&<\mathcal{O}\left[z^{D-1}\int_{0}^{2}dy\,y^{D-2}e^{-\frac{\kappa z}{a_{0}}\left(1+\frac{y^{2}}{4}\right)}\right]\nonumber\\
&\quad+\mathcal{O}\left[z^{D-1}\int_{2}^{\infty}dy\,y^{D-2}e^{-\frac{\kappa z}{a_{0}}y}\right].
\end{eqnarray}
The first term is less than
\begin{equation}
\mathcal{O}\left[z^{D-1}e^{-\frac{\kappa}{a_{0}}z}\int_{0}^{\infty}dy\,y^{D-2}e^{-\frac{\kappa z}{4a_{0}}y^{2}}\right]
=\mathcal{O}\left[z^{\frac{1}{2}(D-1)}e^{-\frac{\kappa}{a_{0}}z}\right]
\end{equation}
and the second term is
$\mathcal{O}[z^{D-2}\exp(-\frac{2\kappa}{a_{0}}z)]$. Note that both of them asymptotically decay exponentially. Consequently, quasi-locally extended states obey the exponential clustering. 

\section{Entanglement Entropies of the Ground State}

\begin{figure}
\begin{tabular}{cc}
\includegraphics[width=0.46\textwidth]{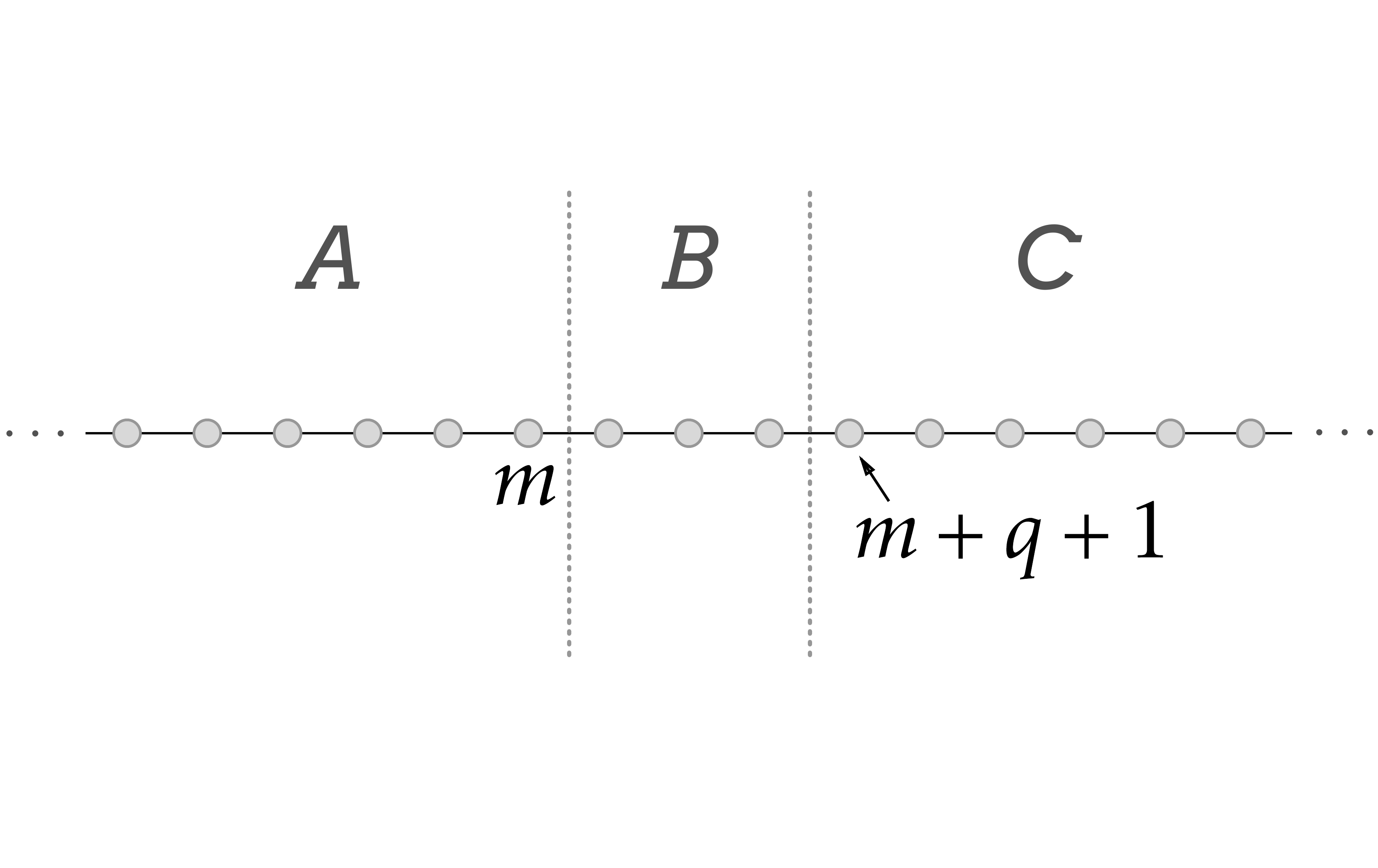} & 
\includegraphics[width=0.46\textwidth]{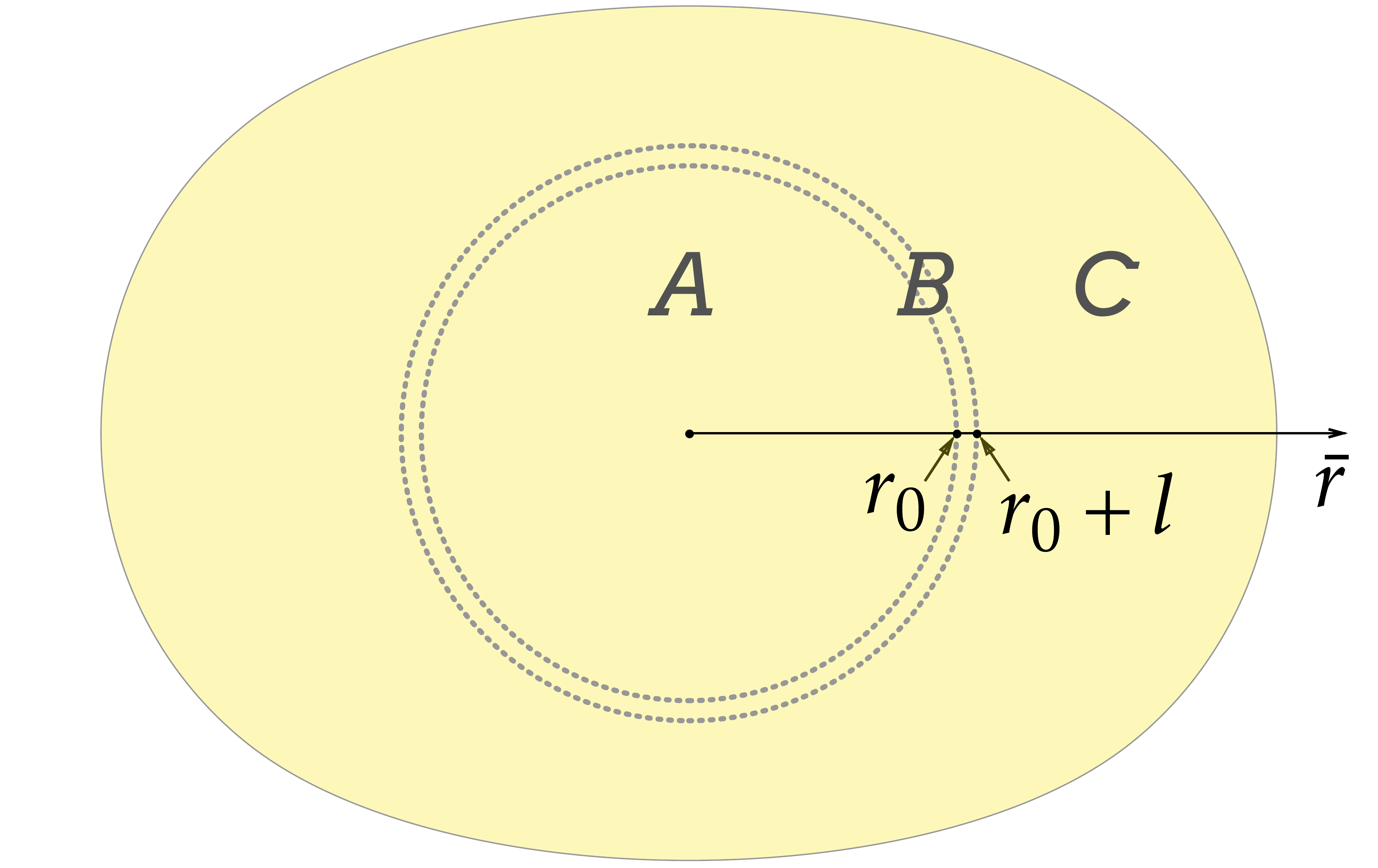} \\
(a) & (b)
\end{tabular}
\caption{Division of the system to bound entanglement entropies (a) in one dimension and (b) in higher dimension.}
\label{fig:alaw}
\end{figure}

The boundary effect function also gives a bound to entanglement entropies of the ground state. Let us first consider a one-dimensional chain of $n$ spins. We divide the chain, as shown in Fig.~\ref{fig:alaw}(a), so that region $A$ is from site 1 to m, $B$ is from $m+1$ to $m+q$, and $C$ is from $m+q+1$ to $n$, where $q>0$ will be chosen later. Our aim is to bound $S(\rho_A\s{n})$. The starting point is the ground state $\sket{\Psi_{0}\s{m+q}}=\sket{\Psi_{0}}_{AB}$ of the $(m+q)$-spin system in region $A+B$. Note that $S(\rho_A\s{m+q})\le q$. Suppose we extend the system by filling up the spins in region $C$. Let $\rho\s{s}=\sket{\Psi_{0}\s{s}}\sbra{\Psi_{0}\s{s}}$. From the variant of Fannes' inequality~\cite{alicki04},
\begin{eqnarray}
\abs{S(\rho_A\s{s})-S(\rho_A\s{s-1})}&=\abs{S(\rho\s{s}|\rho_A\s{s})-S(\rho\s{s-1}|\rho_A\s{s-1})}\nonumber\\
&\le 8\eta_A\s{s}(s-m)-2\mathcal{H}_{2}(2\eta_A\s{s}),
\end{eqnarray} 
where $\mathcal{H}_{2}(p)=-p\log p-(1-p)\log(1-p)$. 
As $\mathcal{H}_{2}(p)<(p/\sqrt{2})^{1/2}$, 
\begin{eqnarray}
\abs{S(\rho_A\s{s})-S(\rho_A\s{s-1})}&<8\eta_A\s{s}(s-m)+2(\sqrt{2}\eta_A\s{s})^{1/2}\nonumber\\
&\le8\sqrt{2}(s-m)\mu(s-m)+2\sqrt{2\mu(s-m)}.
\end{eqnarray}
It follows that
\begin{eqnarray}
S(\rho_A\s{n})&\le S(\rho_A\s{m+q})+\sum_{s=m+q+1}^{n}\abs{S(\rho_A\s{s})-S(\rho_A\s{s-1})}\nonumber\\
&\lesssim q+2\sqrt{2}\int_{q}^{\infty} dr\{4r\mu(r)+\sqrt{\mu(r)}\}.
\end{eqnarray}
For any constant $q$, $S(\rho_A\s{n})$ is bounded by a constant, i.e., the entanglement area law is obeyed, as far as $\mu(r)$ asymptotically decreases faster than $r^{-2}$.

In higher dimension, we divide the system, as shown in Fig.~\ref{fig:alaw}(b), so that regions $A$, $B$, and $C$ are defined by $\bar{r}\le r_{0}$, $r_{0}<\bar{r}\le r_{0}+l$, and $\bar{r}>r_{0}+l$, respectively, where $\bar{r}$ is the Euclidean distance from the origin and $l\ll r_{0}$ will be chosen later. Suppose there are $m$, $q$, and $n-m-q$ spins in regions $A$, $B$, and $C$, respectively. We again start from $\sket{\Psi_{0}\s{m+q}}=\sket{\Psi_0}_{{AB}}$ and extend it by filling up the spins in region $C$ in such a way that the shape of the system is kept almost the same but only the size is gradually increased. In this way, the integral is simplified. Note that
\begin{equation}
q\le\mathcal{O}[n_{0}\{(r_{0}+l)^{D}-r_{0}^{D}\}]=\mathcal{O}(lr_{0}^{D-1}).
\end{equation} 
In the same way as above,
\begin{eqnarray}
S(\rho_A\s{n})\le& S(\rho_A\s{m+q})\nonumber\\
&+\mathcal{O}\left[\int_{l}^{\infty}dy\,(r_{0}+y)^{D-1}n_{0}\{(r_{0}+y)^{D}-r_{0}^{D}\}\mu(y/a_{0})\right]\nonumber\\
&+\mathcal{O}\left[\int_{l}^{\infty}dy\,(r_{0}+y)^{D-1}\sqrt{\mu(y/a_{0})}\right].
\end{eqnarray} 
If $\mu(r)\le\mu_{0}e^{-\kappa r}$,
\begin{eqnarray}
S(\rho_A\s{n})\le&\mathcal{O}(lr_{0}^{D-1})+\mathcal{O}\left[(r_{0}+l)^{2D-2}l\exp\left(-\frac{\kappa}{a_{0}}l\right)\right]\nonumber\\
&+\mathcal{O}\left[(r_{0}+l)^{D-1}\exp\left(-\frac{\kappa}{2a_{0}}l\right)\right].
\end{eqnarray} 
By taking $l=\frac{a_{0}}{\kappa}(2D-2)\log r_{0}$, we have
\begin{equation}
S(\rho_A\s{n})\le\mathcal{O}(r_{0}^{D-1}\log r_{0}),
\end{equation} 
i.e., a quasi-locally extended state obeys the entanglement area law up to a logarithmic correction for $D\ge 2$. Note that this bound has been derived solely from the exponentially decaying boundary effect function, regardless of the gapness of the system. For gapped systems, one can take an alternative approach to eliminate the logarithmic correction~\cite{cho14}.

\section{Conclusion}

Summing up, the boundary effect function has been introduced as a means to characterise the convergence of the ground state towards its thermodynamic limit and the degree to which the boundary effect permeates into the bulk. It has been proven that the boundary effect function also bounds two-point correlation functions and entanglement entropies of the ground state. In particular, for gapped local Hamiltonians with a nondegenerate ground state, the boundary effect function decays exponentially except for an unusual exception, implying that the ground state is quasi-locally extended. It turned out that such a local nature of the ground state is manifested as restricted correlations in the ground state. In particular, quasi-locally extended states obey the exponential clustering and the entanglement area law up to a logarithmic correction in any spatial dimension.  As a final remark, we would like to emphasise that the boundary effect function is defined on its own, independently from other attributes of the system. For example, while an exponentially decaying boundary effect function implies restricted correlations, there is no necessary reason to believe that it implies a finite spectral gap as a finite correlation length does not generally imply a finite spectral gap~\cite{nac96}. Furthermore, in one dimension, the area law can be obeyed even for a moderately decaying boundary effect function, in which case the system is thermodynamically gapless. Our work would thus complement the earlier results on the ground state correlations in local many-body systems, e.g., those in gapped systems~\cite{nach06,hastings06,hastings07,cho14}, thereby deepening and enriching our understanding of them.

\ack

The author thanks M.-J. Hwang, H.-J. Kim, and S.-W. Ji for discussions.

\section*{References}

\bibliography{references}

\providecommand{\newblock}{}
\begin{thebibliography}{10}
\expandafter\ifx\csname url\endcsname\relax
  \def\url#1{{\tt #1}}\fi
\expandafter\ifx\csname urlprefix\endcsname\relax\def\urlprefix{URL }\fi
\providecommand{\eprint}[2][]{\url{#2}}

\bibitem{lieb72}
Lieb E~H and Robinson D~W 1972 {\em Commun. Math. Phys.\/} {\bf 28} 251

\bibitem{nach06}
Nachtergaele B and Sims R 2006 {\em Commun. Math. Phys.\/} {\bf 265} 119

\bibitem{hastings06}
Hastings M~B and Koma T 2006 {\em Commun. Math. Phys.\/} {\bf 265} 781

\bibitem{eisert10}
Eisert J, Cramer M and Plenio M~B 2010 {\em Rev. Mod. Phys.\/} {\bf 82} 277

\bibitem{hastings07}
Hastings M~B 2007 J. Stat. Mech. P08024

\bibitem{page93}
Page D~N 1993 {\em Phys. Rev. Lett.\/} {\bf 71} 1291

\bibitem{scho11}
Schollw\"ock U 2011 {\em Ann. Phys.\/} {\bf 326} 96

\bibitem{kitaev06}
Kitaev A and Preskill J 2006 {\em Phys. Rev. Lett.\/} {\bf 96} 110404

\bibitem{levin06}
Levin M and Wen X~G 2006 {\em Phys. Rev. Lett.\/} {\bf 96} 110405

\bibitem{gha14}
Gharibian S, Huang Y, Landau Z and Shin S~W 2014 arXiv:1401.3916

\bibitem{aud02}
Audenaert K, Eisert J, Plenio M~B and Werner R~F 2002 {\em Phys. Rev. A\/} {\bf
  66} 042327

\bibitem{plenio05}
Plenio M~B, Eisert J, Drei{\ss}ig J and Cramer M 2005 {\em Phys. Rev. Lett.\/}
  {\bf 94} 060503

\bibitem{wolf06}
Wolf M~M 2006 {\em Phys. Rev. Lett.\/} {\bf 96} 010404

\bibitem{wolf08}
Wolf M~M, Verstraete F, Hastings M~B and Cirac J~I 2008 {\em Phys. Rev.
  Lett.\/} {\bf 100} 070502

\bibitem{masanes09}
Masanes L 2009 {\em Phys. Rev. A\/} {\bf 80} 052104

\bibitem{beau10}
de~Beaudrap N, Ohliger M, Osborne T~J and Eisert J 2010 {\em Phys. Rev.
  Lett.\/} {\bf 105} 060504

\bibitem{gott10}
Gottesman D and Hastings M~B 2010 {\em New J. Phys.\/} {\bf 12} 025002

\bibitem{aharonov11}
Aharonov D, Arad I, Vazirani U and Landau Z 2011 {\em New J. Phys.\/} {\bf 13}
  113043

\bibitem{arad12}
Arad I, Landau Z and Vazirani U 2012 {\em Phys. Rev. B\/} {\bf 85} 195145

\bibitem{mic12}
Michalakis S 2006 arXiv:1206.6900

\bibitem{mic13}
Michalakis S and Zwolak J~P 2013 {\em Commun. Math. Phys.\/} {\bf 322} 277

\bibitem{aco13}
Van~Acoleyen K, Mari\"en M and Verstraete F 2013 {\em Phys. Rev. Lett.\/} {\bf
  111} 170501

\bibitem{bra13}
Brand{\~a}o F~G~S~L and Horodecki M 2013 {\em Nature Phys.\/} {\bf 9} 721

\bibitem{cho14}
Cho J 2014 {\em Phys. Rev. Lett.\/} {\bf 113} 197204

\bibitem{skl88}
Sklyanin E~K 1988 {\em J. Phys. A: Math. Gen.\/} {\bf 21} 2375

\bibitem{gho94}
Ghoshal S and Zamolodchikov A 1994 {\em Int. J. Mod. Phys. A\/} {\bf 9} 3841

\bibitem{wen91}
Wen X~G 1991 {\em Phys. Rev. B\/} {\bf 43} 11025

\bibitem{has10}
Hasan M~Z and Kane C~L 2010 {\em Rev. Mod. Phys.\/} {\bf 82} 3045

\bibitem{nielsen}
Nielsen M and Chuang I 2000 {\em Quantum Computation and Quantum Information\/}
  (Cambridge University Press, Cambridge, England)

\bibitem{lidar09}
Lidar D~A, Rezakhani A~T and Hamma A 2009 {\em J. Math. Phys.\/} {\bf 50}
  102106

\bibitem{lidar08}
Lidar D~A, Zanardi P and Khodjasteh K 2008 {\em Phys. Rev. A\/} {\bf 78} 012308

\bibitem{alicki04}
Alicki R and Fannes M 2004 {\em J. Phys. A\/} {\bf 37} L55

\bibitem{nac96}
Nachtergaele B 1996 {\em Commun. Math. Phys.\/} {\bf 175} 565

\end{thebibliography}

\end{document}